\documentclass{aa}
\usepackage{graphicx}
\def \ha   {H$\alpha$}
\def \oiii {[O~{\sc iii}]~5007~\AA}

\def \NII  {[N~{\sc ii}]~6548~\&~6584~\AA}

\def \tena#1 #2 {\ifmmode{#1 \times 10^{#2}}\else{$#1 \times 10^{#2}$}\fi}
\def \kms  {\ifmmode{~{\rm km\,s}^{-1}}\else{~km s$^{-1}$}\fi}
\def \vhel {\ifmmode{V_{{\rm HEL}}}\else{~$V_{{\rm HEL}}$}\fi}
\def \vsys {\ifmmode{V_{{\rm sys}}}\else{~$V_{{\rm sys}}$}\fi}
\def \vexp {\ifmmode{V_{{\rm exp}}}\else{~$V_{{\rm exp}}$}\fi}

\def \deg  {\ifmmode{^{\circ}}\else{$^{\circ}$}\fi} 
\def \msun {\ifmmode{{\rm\ M}_\odot}\else{${\rm\ M}_\odot$}\fi}
\def \myr  {\ifmmode{{\rm\ M}_\odot{\rm\ yr}^{-1}}
         \else{${\rm\ M}_\odot$ yr$^{-1}$}\fi}
\def\mnras{MNRAS}
\def\apj{ApJ}
\def\apjs{ApJS}
\def\aap{A\&A}

\begin{document}

\title{The kinematics of the bi-lobal supernova remnant G~65.3+5.7 -- 
Paper II}

\subtitle{}

\author{P. Boumis\inst{1} 
     \and   J. Meaburn\inst{2}
     \and   J. A. L{\'o}pez\inst{3}  
     \and   F. Mavromatakis\inst{4}
     \and   M. P. Redman\inst{5}
     \and   D.J. Harman\inst{2}
     \and C. D. Goudis\inst{1,6}
          }

\offprints{Dr. P. Boumis}
\authorrunning{P. Boumis et al.}
\titlerunning{The kinematics of G 65.3+5.7}
\institute{Institute of Astronomy \& Astrophysics, National Observatory of
Athens, I. Metaxa \& V. Paulou, P. Penteli, GR-15236 Athens, Greece.
\and Jodrell Bank Observatory, University of Manchester, Macclesfield 
SK11 9DL, UK.
\and Instituto de Astronomia, UNAM, Apdo. Postal 877. Ensenada, B.C. 22800,
M\'{e}xico.
\and University of Crete, Physics Department, P.O. Box 2208, 710 03 
Heraklion, Crete, Greece.
\and Dublin Institute for Advanced Studies, School of Cosmic Physics, 
5 Merrion Square, Dublin 2, Republic of Ireland.
\and Astronomical Laboratory, Department of Physics, University 
of Patras, 26500 Rio--Patras, Greece.
}

\date{Received 9 March 2004 / Accepted 11 May 2004}

\abstract{Further deep, narrow-band images in the light of \oiii\ have
been added to the previous mosaic of the faint galactic supernova
remnant G~65.3+5.7. Additionally longslit spatially resolved
\oiii\ line profiles have been obtained at sample positions using the 
Manchester Echelle Spectrometer at the San Pedro Martir observatory.
The remnant is shown to be predominantly bi-lobal with an EW axis for
this structure. However, a faint additional northern lobe has now been
revealed.

Splitting of the profiles along the slit lengths, when extrapolated to
the remnant's centre, although uncertain suggests that the expansion
velocity of this remnant is between 124 and 187 \kms\ ie much lower
than the 400 \kms\ previously predicted for the forward shock velocity
from the X--ray emission.

An expansion proper motion measurement of 2.1 $\pm$ 0.4 arcsec in 48
years for the remnant's filamentary edge in the light of \ha\ + \NII\
has also been made. When combined with an expansion velocity of
$\approx$ 155 \kms, a distance of $\approx$ 800 pc to G~65.3+5.7 is
derived.

Several possibilities are considered for the large difference in the
expansion velocity measured here and the 400 \kms\ shock velocity
required to generate the X--ray emission. It is also suggested that
the morphology of the remnant may be created by a tilt in the galactic
magnetic field in this vicinity.
 
\keywords{ISM:general--ISM: supernova remnants--ISM: individual objects:
 G~65.3+5.7}}

\maketitle

\section{Introduction}

Mavromatakis et al. (2002 - hereafter called Paper I) presented the
deepest optical images of the supernova remnant (SNR) G~65.3+5.7,
discovered by Gull et al. (1977), yet obtained. They revealed a
filamentary structure $\approx$ 3\deg\ $\times$ 4\deg\ in size,
emitting the \oiii\ nebular line brightly and predominantly composed
of two irregular filamentary rings with displaced centres; a bi-lobal
morphology is implied.

Also presented in Paper I are low--dispersion spectra at eight
positions (P1--8) around the remnant's perimeter; these are consistent
with ionization by a 90--140 \kms\ shock with a local ionized gas
density of 200 cm$^{-3}$ typical of an $\approx$ 20,000 yr old
SNR. {\it ROSAT} observations by Aschenbach (1994) and Lu \&
Aschenbach (2004) reveal clumpy X--ray emission from G~65.3+5.7,
assuming a distance of 1 kpc, from a forward shock of $\approx$ 400
\kms\ in a tenuous ambient medium of density 0.02 cm$^{-3}$.

Losinskaya (1980), from \ha\ interferograms obtained with a classical
Fabry--Perot interferometer, had extrapolated line splitting over the
bright filamentary edge of G~65.3+5.7 (called `A newly discovered SNR
in Cygnus' in that paper) to give a global expansion velocity
\vexp\ = 400 $\pm$ 200 \kms.

Deep, spatially resolved profiles of the \oiii\ emission line have now
been obtained with an echelle spectrometer at positions P1--5 \& 7--8
where the Paper I low--dispersion spectra were obtained, and at a
further position, P9, over the centre of G~65.3+5.7 with the intention
of measuring the maximum expansion velocity directly as well as
obtaining accurate line profiles at representive positions. Further
deep CCD images, again in the \oiii\ line and neighbouring continuum,
were also obtained, with the same wide--field imaging system employed
in Paper I; a more complete continuum subtracted image of this SNR has
resulted.

\section{Observations and results}

\subsection{New imagery}

Three further \oiii\ and continuum images were obtained on June
2003 27-28 with the 89\arcmin\ $\times$ 89\arcmin\ field (5\arcsec\ per
pixel) 0.3 m Schmidt Cassegrain telescope at the Skinakas Observatory,
Crete, Greece to complete the coverage of the westerly edge of the
SNR.  These were then combined, after an identical analysis, with the
\oiii\ image described in Paper I. Light and deep negative grey-scale
representations of the whole field are shown in Figs. 1a-b
respectively.

\subsection{Long-slit spectroscopy}

The present spectral observations were made with the Manchester
Echelle Spectrometer (MES-SPM - see Meaburn et al. 1984; 2003)
combined with the 2.1-m San Pedro Martir telescope. A SITe3 CCD was
the detector with 1024$\times$1024, 24~$\mu$m pixels although
2$\times$2 binning was employed throughout the observations on the
nights of the May 2003 29-31.

Spatially resolved, long-slit line profiles at high spectral
resolution were obtained with the MES--SPM. This spectrometer has no
cross--dispersion. For the present observations, a filter of 60~\AA\
bandwidth was used to isolate the 114$^{\rm th}$ echelle order
containing the \oiii\ nebular emission line.

The 512 increments, each 0.626\arcsec\ long, give a total projected
slit length of 5\farcm34 on the sky. `Seeing' was always $\leq
1$\arcsec\ during these observations.  A 150~$\mu$m wide ($\equiv
12$~$\kms$ and 1.9\arcsec) single slit was used.

The data were bias corrected, cleaned etc. in the usual way using the
STARLINK \textsc{figaro} and \textsc{kappa} software packages. All
spectra were calibrated in heliocentric radial velocity (\vhel) to
$\pm$~3~$\kms$ accuracy against spectra of a thorium/argon
lamp. Absolute surface brightnesses, B$_{\rm [O~III]}$ erg s$^{-1}$
cm$^{-2}$ sr$^{-1}$ \AA$^{-1}$, of the line profiles were obtained by
comparing the spectra to the slitless spectrum of the standard star
Feige 56. All spectra were obtained in photometric conditions, and
without correction for interstellar extinction are accurate to around
$\pm$ 10 percent.

The slit was orientated EW on the centres P1-5 \& P7--8 from Paper I
with integration times of 1800s. The integration time for a further
slit position (P9) was 3600s. These slit positions are shown against
the lightly printed image of the SNR in Fig. 1a.

Negative grey-scale representations of the position--velocity (pv)
arrays of \oiii\ line profiles are shown in Figs~2a--h for slit
positions P1--5 \& P7--8. More details of the imagery and pv arrays
over the brighter filaments are shown in Figs. 3a--b and 4a--b
for slits P1 and P2 respectively. 

The line profiles extracted from various lengths of the pv arrays in
Fig. 2a--h and for P9, where the emission is very faint, are shown in
Fig. 5. The observed peak value of B$_{\rm [O~III]}$ in units of
10$^{-6}$ erg s$^{-1}$ cm$^{-2}$ sr$^{-1}$ \AA$^{-1}$ is given next to
each profile.

\subsection{Expansion Proper Motions}

The filament at Pos. 2 (Fig.~1a) was detected by the Palomar
Observatory Sky Survey POSS--E red plate taken in 1951.  This is 48 yr
before the \ha\ + \NII\ image was obtained (taken in 1999 July 13
similarly to those in Fig. 1a \& b -- see Paper I for details). Both
images are predominantly in the light of the \ha\ + \NII\ lines and
the Digitized Sky Survey of the POSS plate has a very similar angular
resolution to the 1999 imagery. A measurement of the expansion proper
motion of G~65.3+5.7 has therefore been made.

Firstly, the two time-separated images were rotated identically until
the filament covered by Pos. 2 became vertical in both
arrays. Profiles were then extracted along identical lines where there
were no confusing stellar images on the filament but many reference
star profiles further away. Gaussians were fitted to both faint
stellar and filament profiles to give a measured shift perpendicular
to the filament's length, and away from the centre of the supernova
remnant, of $\delta\theta_{pm}$ = 2.1 $\pm$ 0.4 arcsec in 48 years.

\section{Discussion}

\subsection{Kinematics and distance}

The first useful parameter to determine for such an extended SNR is
its systemic heliocentric radial velocity (\vsys). This is given most
plausibly by the mean \vhel\ of the profiles of the brightest
filaments on the perimeter of the remnant for it is assumed that these
are being viewed tangentially. The centroids of the single profiles in
Figs. 3, 4 \& 5 for the filamentary edges along slits P1, 2, 3, 5 and
7 are at \vhel\ = -19, 0, 0, -10 and -15 \kms\ respectively to give a
value of \vsys\ = -7 \kms\ for G~65.3+5.7 (i.e. halfway between the
mean radial velocities of the northern and southern filaments). Note
that the profiles for the northern filaments (P1, P7 and P5) are
significantly displaced to approaching radial velocities compared with
their southern counterparts (P2 \& P3).  The halfwidths of the same
profiles (for P1, 2, 3, 5 \& 7) fitted by single Gaussians and
corrected for the instrumental broadening are 0.42 $\pm$ 0.02, 0.44
$\pm$ 0.03, 0.47 $\pm$ 0.01, 0.46 $\pm$ 0.08 \& 0.36 $\pm$ 0.01 \AA\
respectively. As the post-shock interstellar gas emitting \oiii\ has
most likely cooled to T$_{e}$ = 10$^{4}$ K then these widths imply
residual turbulent motions, combined with non-turbulent motions due to
curvature of the shock fronts, of around 23 \kms\ in all of these
filamentary edges for the width of the thermal component of the \oiii\
line will only be 0.09 \AA\ as given by $\delta\lambda$ = 8.927
$\times$ 10$^{-4}$ $\times$ T$_{e}^{1/2}$ \AA.

Knowledge of the global expansion velocity \vexp\ for such a complex
remnant is also desirable. In the pv arrays in Fig. 2, as detected by
Lozinskaya (1980), there is evidence of expansion along slits P2 and
P7: the profiles are split by about 70 \kms\ away from the bright
filamentary edge, towards the westerly end of P2 (and see this trend
along the short length shown in Fig. 4b) and go from -15 to -60 \kms\
along the length of P7.  If it is assumed that this splitting towards
the westerly end of P2 is due to the spherical expansion of the
91\arcmin\ radius easterly lobe of G~65.3+5.7 then, after taking
account of the slit orientation with respect to the filamentary edge,
by extrapolation \vexp\ = 187 $\pm$ 50 \kms\ is derived for this lobe
of the remnant. The lower end of the \vexp\ values given by Losinskaya
(1980) ( to $\pm$ 200 \kms\ accuracy) is therefore favoured.

Other clues to the true value of \vexp\ are also in the present
observations. For instance velocity components further from \vsys\ are
found along slit positions P8 and 9 which are well inside the
remnant's outer perimeter (Fig. 1a). In the pv array in Fig. 2 for P8
a component at \vhel\ $\approx$ 75 \kms\ can be seen. If this reflects
the expansion of the easterly lobe in Fig. 1a, and assuming
\vsys\ = -7 \kms\ and 
spherical expansion, then \vexp\ $\approx$ 124 \kms\ for the lobe for
this position is $\approx$ 0.25 of the remnant's radius towards the
lobe's centre; whereas the component with \vhel\ = 0 \kms\ along the
same slit length must be from the filamentary edge of the western lobe
(see Fig. 1 a).  Similarly, the faint profile from P9 in Fig. 5 from
near the center of the eastern lobe but just inside the edge of the
western lobe of the remnant has a velocity component at \vhel\ = +130
\kms\ which when compared with \vsys\ = -7 \kms\ could imply \vexp\
$\approx$ 140 \kms\ maybe for the eastern lobe and most likely for the
whole remnant.
 
All of the current estimations of \vexp\ (between 124 to 187 \kms) are
well short of the value of the 400 \kms\ forward shock velocity
required for the X-ray emission (Lu \& Aschenbach 2004) and below the
bottom end (\vexp\ = 200 \kms) of the large range given by Losinskaya
(1980). If spherical expansion is assumed at \vexp(\kms) in T (yr) to
give $\delta\theta_{pm}$ (arcsec) then the distance D(pc) is given by
D = 0.2168 $\times$ \vexp\ $\times$ T $\times$
$\delta\theta_{pm}^{-1}$ in which case for $\delta\theta_{pm}$ = 2.1
arcsec in 48 yr then D/\vexp\ = 4.955. This gives D = 770 $\pm$ 200 pc
for \vexp\ $\approx$ 155 \kms\ but a large 1980 pc for 400 \kms.

\subsection{Energetics}

As discussed above the optically measured expansion and shock
velocities and the X-ray derived shock velocities differ sharply. This
is also observed for other similar SNR, particularly the Cygnus
Loop. Lu \& Aschenbach (2004) and others have assumed that G65.3+5.7 is
still in its adiabatic phase of evolution, thus enabling a Sedov
analysis of the energetics. The low ambient density derived from this
analysis suggests a lower than usual post-shock cooling rate ($\propto
n^2$) and thus make it less than likely that the SNR will have moved
into the momentum conserving phase. The simplest solution to the
discrepancy is that the central regions were indeed subject to a
$\sim 400~{\rm km~s^{-1}}$ shock in a rarified medium which produced
the X-ray emission but that the shock speed has recently encountered
much denser material which has reduced the shock speed and led to the
optical emission that is observed. Two possible sources of this dense
material are a cloudy ISM and a pre-existing dense circumstellar shell
and these are now discussed in turn.

Lu \& Aschenbach (2004) favour a shock wave propagating into a cloudy
ISM as described in detail by McKee \& Cowie (1975). The two shock
speeds are then due to cloud shocks and the main shock travelling
through the intercloud medium. Less obviously explained is the fact
that the measured expansion velocity of the SNR as measured by the
optical filaments is also much less than the X-ray shock velocity. The
X-ray emission does not extend measurably beyond the boundary
delimited by the optical filaments. This requires that the clouds are
promptly disrupted by the shock, incorporated into the postshock flow
and that recombination takes place. For small clouds, the disruption
takes a few `cloud crushing' times, as defined by Klein, McKee and
Colella (1994), by which time the leading shock will be a few times
$\chi^{1/2}$ ahead of the cloud, where $\chi$ is the density contrast
between the cloud and intercloud medium. For a $\chi < 100$ this leads
to a displacement of a few tens of the cloud size. As long as the
density contrast or cloud sizes are not too large the optically
emitting gas will be close to the boundary of the SNR and moving with
a velocity a fraction of the initial shock speed.
The difficulty with such a picture is that a very large number of small
clouds are required to get the very smooth O~{\sc iii} distribution as
seen in our figures.

Note that the X-ray shock velocity estimate comes from the current
X-ray temperature and if this has been cooled significantly by the
addition of cold stationary clouds then the actual X-ray shock
velocity should be even higher. However, McKee \& Cowie (1975) argue
that the acceleration and evaporation of the over-run clouds will have
little effect on the energetics of the SNR as long as their filling
factor is not too large. Some support for this model comes from the
clumpy nature of the X-ray emission as mapped by Lu \& Aschenbach
(2004).
 
An alternative is that a pre-existing dense circumstellar shell has been
encountered by the blast wave and this has led to the two different shock
velocities. If the progenitor was a massive star then an
H~{\sc ii} region and wind blown shell will have been set up around
the star. Shull et al. (1985) discussed the effects of a SN explosion
in a cloudy medium around a massive star. The formation of the H~{\sc
ii} region destroys any H~{\sc i} clouds and thus the SN explosion
takes place in a cavity created by the star and then encounters a
clumpy shell. Similarly, Charles, Kahn \& McKee (1985) argue that the
Cygnus loop SNR was caused by a SNR exploding in a pre-existing
cavity, presumably generated by the progenitor. Such a model has the
attraction that it accounts for the relatively low ambient medium
density implied by a Sedov analysis of the X-ray shock emission. It
also offers a possible explanation for the morphology of G65.3+5.7,
discussed further below, since an axisymmetric mass-loss rate prior to
the SN explosion could later govern the evolution of the shape of the
remnant.

The local electron densities in the recombination zone, as measured in
Paper 1, ranged between $30-170~{\rm cm^{-3}}$. For a temperature of
$10^{4}$ the thermal pressure is then $nT\sim 3\times 10^5~{\rm K~cm^{-3}}
- 2\times 10^6~{\rm K~cm^{-3}}$. In contrast, the thermal pressure of the
X-ray gas implied by the shock speed and density measured by Lu \&
Aschenbach (2004) is only $1.6\times 10^5~{\rm K~cm^{-3}}$. This is
another X-ray/optical discrepancy that has been observed before: Raymond
et al (1988) found a similar effect in the Cygnus loop SNR. There, the
shock ram pressure exceeds the thermal pressure in both the optical and
X-ray emitting gas. Raymond et al (1988) discuss a possibility that could
be relevant here. If a blast wave impacts a large enough cloud, a reverse
shock is generated that halts the expansion of the rarefied X-ray emitting
gas (McKee \& Cowie 1975). This produces a temporary overpressure in the
remnant that then accelerates the shock. The obstacle that causes this
sequence of events could be a previously generated wind blown shell as
discussed above.

\subsection{Morphology}

The more extended, deep image in Fig. 1b reveals clearly the prominent,
nearly circular, filamentary structure, $\approx$ 3\deg\ diameter, 
that must define the edge  of the eastern lobe
which is most probably expanding at $\approx$ 155 \kms.
The prominent but less well--defined 
westerly lobe appears to be composed of at least two
substructures and to complicate matters further a 
fainter northern 
lobe outside of the main perimeter of the remnant can now be seen.
The northern perimeter of the easterly lobe is approaching the observer
at around 16 \kms\ with respect to its southern counterpart. All that can
be assumed is that the two edges have formed in separate interstellar
clouds with different radial velocities.
 
G65.3+5.7 appears to be  a very good example of a barrel--shaped 
supernova remnant (see
Kestevan \& Caswell 1987 for a full discussion and many
examples). There is a clear axis of symmetry, approximately east-west,
either side of which are two bright limbs of emission. Such a
morphology is common and the mechanisms for producing it can be
divided into extrinsic and intrinsic effects (Gaensler 1998). An
example of the former is a well ordered local ISM magnetic field that
is then incorporated into the shell of the SNR. An example of the
latter is a previous axisymmetric outflow from the progenitor that
then influences the evolution of the SNR (see above). Gaensler (1998)
presents evidence that the barrel axes tend to be aligned with the
galactic plane which then suggests an extrinsic effect (specifically,
the galactic magnetic field stratifying the ambient medium) is
responsible in many cases. The barrel axis of G65.3+5.7 is at approximately
45 degrees to the plane and the remnant is displaced from the plane by
$\sim 100~{\rm pc}$. This is of the order of the gas scale height and
so the orientation of the field could be begining to depart from being
parallel to the plane. Though the exact orientation is unclear, 45
degrees is not unreasonable. It does seem unlikely however that
interstellar clouds will be flattened and stratified at this angle to
the ISM. One of the several proposed mechanisms involving the
compression of magnetic field (discussed by Gaensler 1998) and the
generation of the bright limbs could be responsible instead.
 
\section{Conclusions}

New images of G65.3+5.7 emphasize that this is a good example
of a barrel--shaped supernova remnant.

The high resolution line profiles presented here suggest that the global
expansion is $\approx$ 155 \kms\ which is around the shock velocity
deduced from previous low--resolution spectra but well short of
the 400 \kms\ forward shock velocity required to generate the X-ray emission.

This expansion velocity combined with an expansion proper motion 
measurement (2.1 $\pm$ 0.4 arcsec in 48 yr) gives a distance of
770 $\pm$ 200 pc.

It would be desirable to derive a more certain expansion velocity
by obtaining further deep line profiles near the remnant's
geometrical centre. The spectral resolution could be relaxed to
$\approx$ 30 \kms\ by broadening the slit width to ensure sufficient
signal to noise ratio in a reasonable observing time for such faint
nebulosity.

Furthermore, a more accurate expansion proper motion could be derived
by obtaining current  higher resolution (say 1\arcsec) images of the 
filamentary edges though a filter
that matches that of the baseline {\it POSS} image.  

\begin{acknowledgements}

We acknowledge the excellent support of the staff at the Skinakas and
SPM observatories during these observations. JAL gratefully acknowledges
financial support from CONACYT (M\'ex) grants 32214-E and 37214 and
DGAPA-UNAM IN114199. Skinakas Observatory is a collaborative project
of the University of Crete, the Foundation for Research and
Technology-Hellas, and the Max-Planck-Institut f\"{u}r
extraterrestrische Physik.
\end{acknowledgements}

\begin{figure*}
\centering
\includegraphics[height=0.9\textheight]{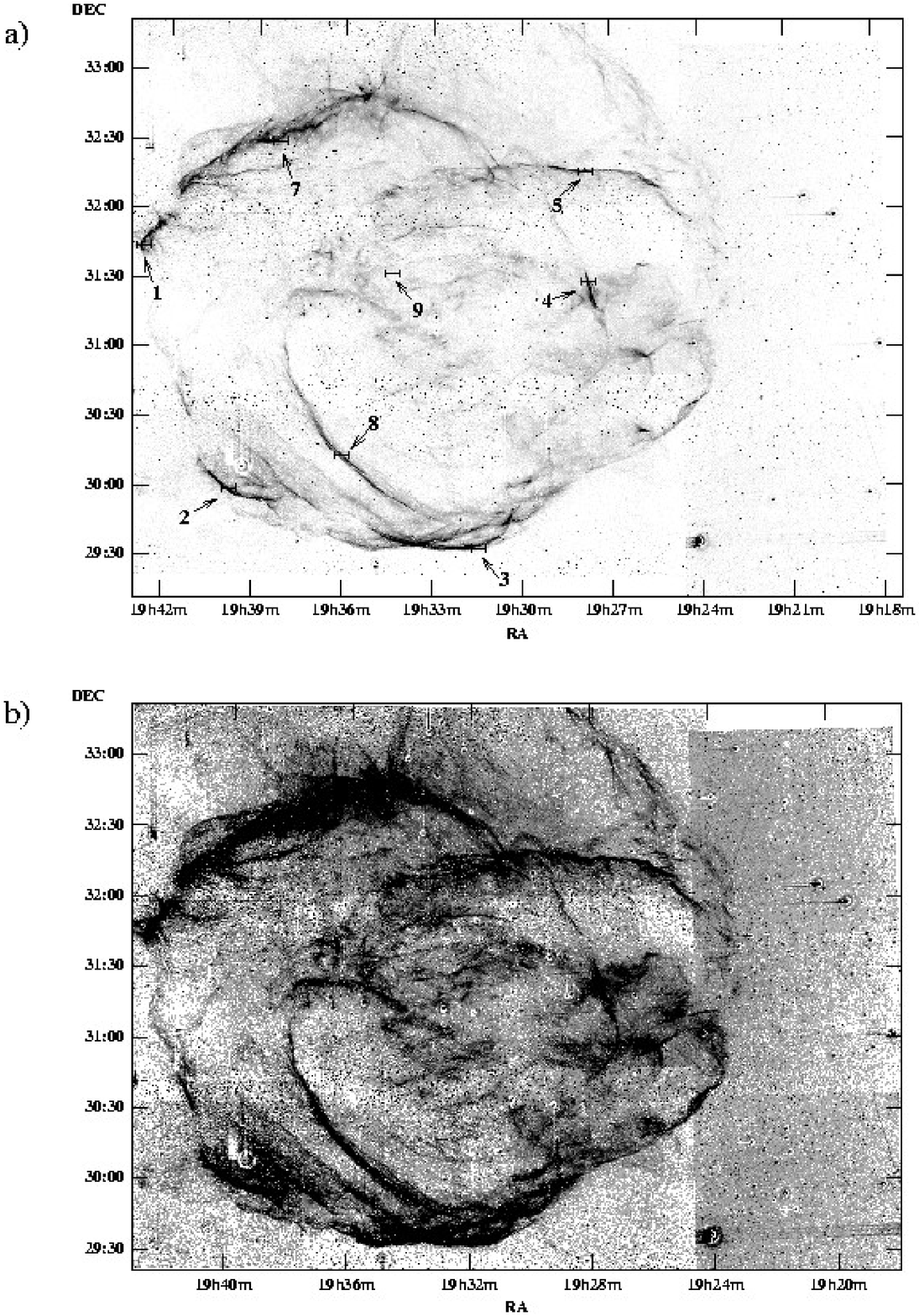}
\caption{A negative, continuum-subtracted mosaic of \oiii\ images
of G~65.3+5.7 is presented  lightly in a) with the slit positions
P1--5 and 7--9 marked and where
the prominent eastern and western lobes are apparent; and  deeply in b) 
to reveal the fainter structure with a further northern lobe suggested
(coords. are 2000 epoch).}
\label{Fig1a-b} 
\end{figure*}

\begin{figure*}
%\centering
\includegraphics[height=0.9\textheight]{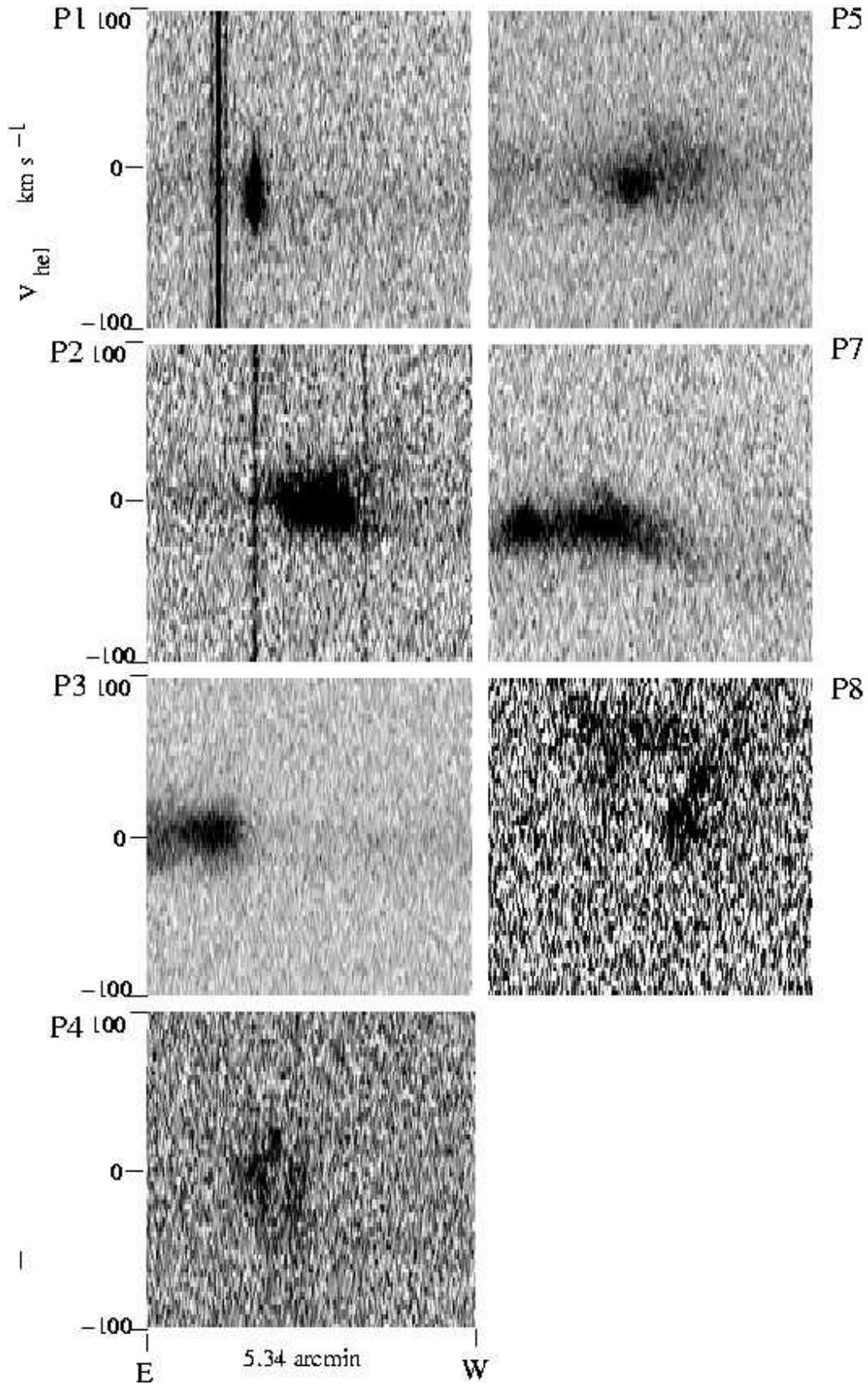}
\caption{Negative greyscale presentations of the \oiii\ position--velocity
arrays
for EW slit positions (See Fig. 1a} P1--5 and 7--8.
\label{Fig2} 
\end{figure*}

\begin{figure*}
%\centering
\includegraphics[height=0.9\textheight]{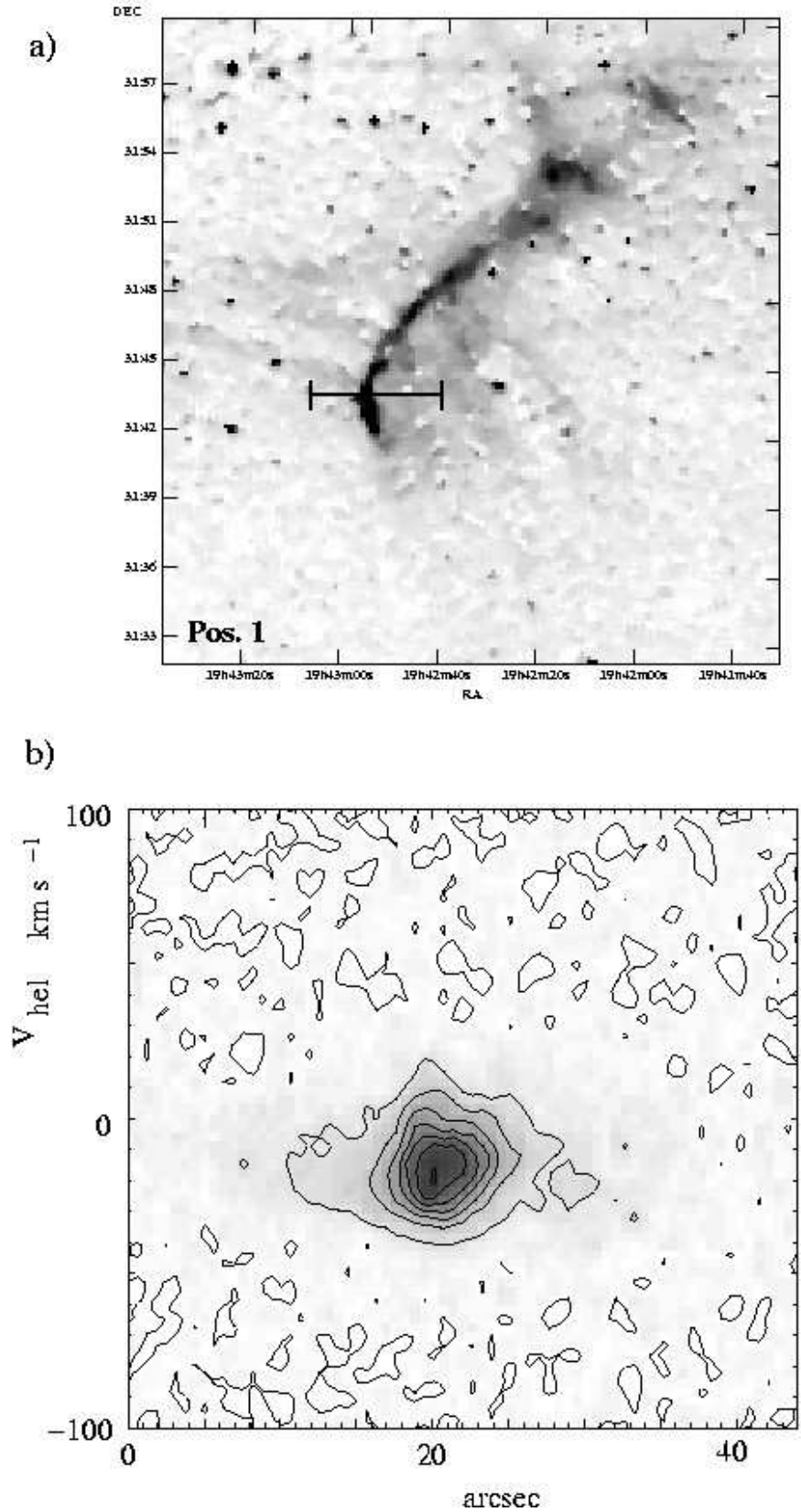}
\caption{a) A small section of the image in Fig.~1 is shown with the
full length of slit P1 marked. b) The contour map is shown, 
with linear intervals,
of a 43\arcsec\ section
of the pv array for P1 (see Fig.~2) where the slit intersects the
bright filament.
Here B$_{\rm [O~III]}$ contours are separated separated by 1.25 $\times$
10$^{-6}$ erg s$^{-1}$ cm$^{-2}$ sr$^{-1}$ \AA$^{-1}$.
} 
\label{Fig3} 
\end{figure*}

\begin{figure*}
%\centering
\includegraphics[height=0.9\textheight]{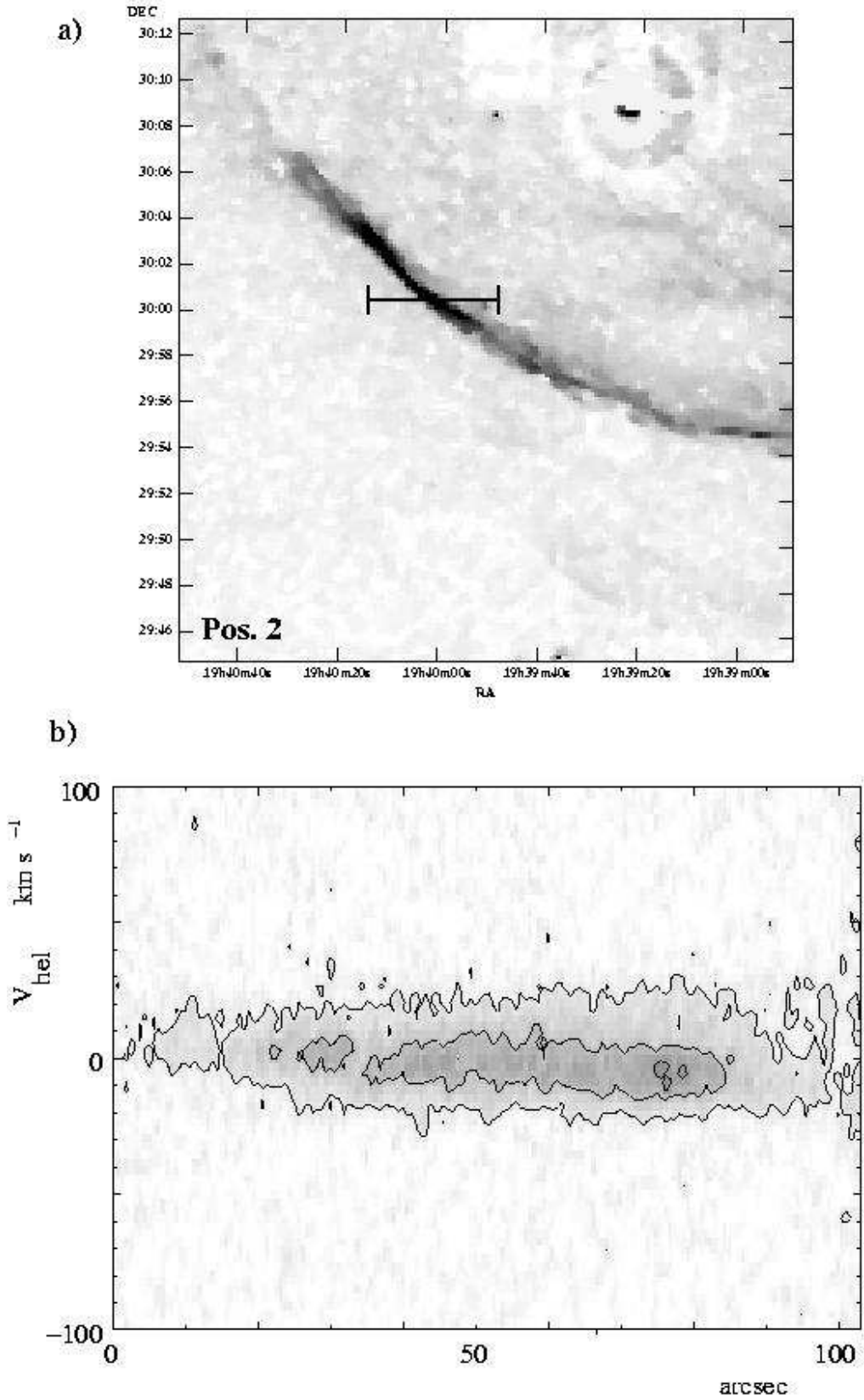}
\caption{As for Fig.~3a--b but for a 102\arcsec\ length of slit P2.
The contour intervals are 0.85 $\times$
10$^{-6}$ erg s$^{-1}$ cm$^{-2}$ sr$^{-1}$ \AA$^{-1}$. }
\label{Fig4} 
\end{figure*}

\begin{figure*}
%\centering
\includegraphics{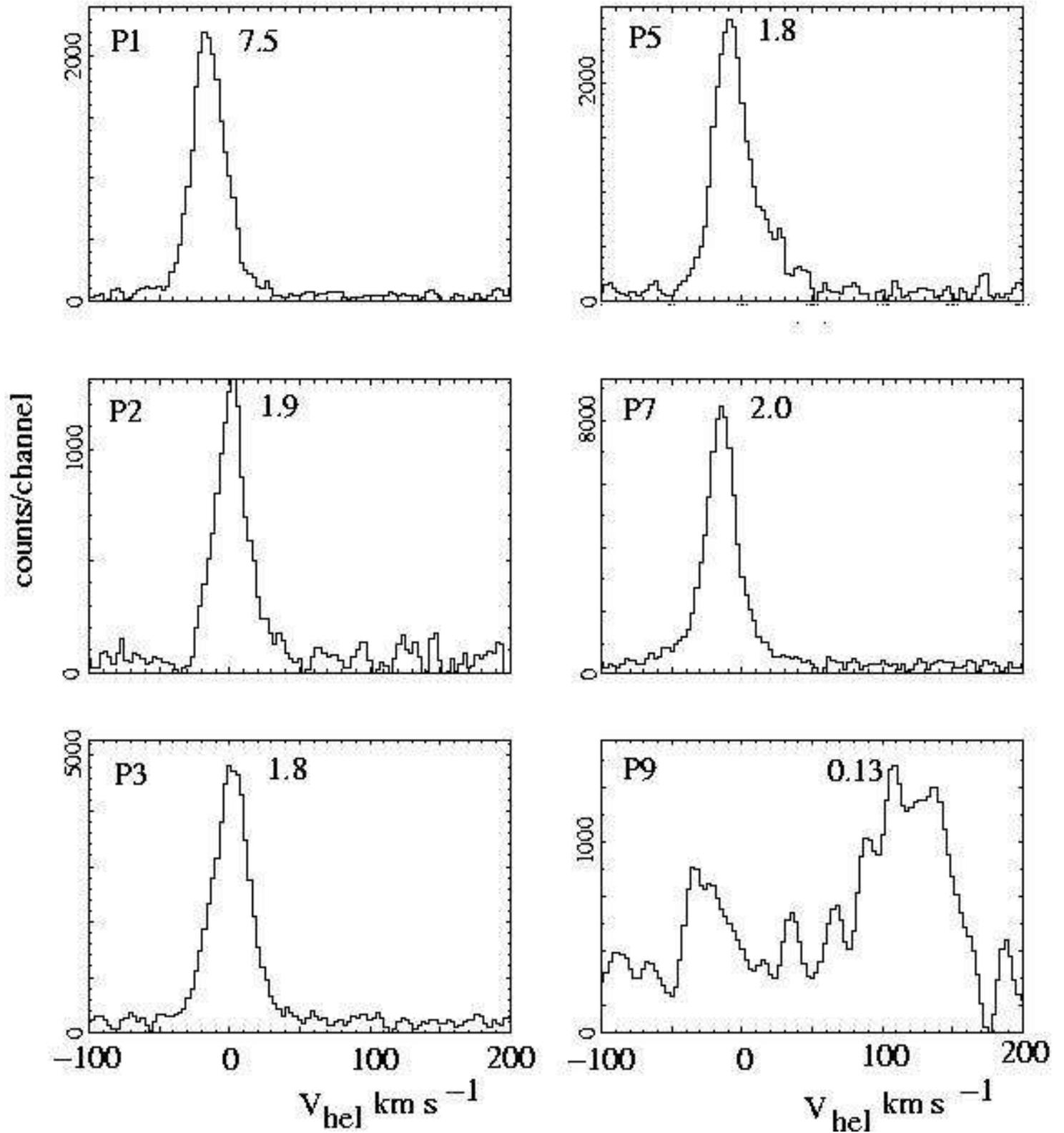}
\caption{
\oiii\ line profiles for the 
brightest filamentary regions intersected by the pv 
arrays in Fig.~2,
and for the diffuse emission along slit P9, 
are presented. The profiles are for the various lengths 
(between the brackets) along the slits: for slit
P1 (7\arcsec), P2 (17\arcsec), P3 (61\arcsec), P5 (33\arcsec), P7 (104\arcsec)
and for P9 (113\arcsec - towards the westerly 
end of the slit position shown in
Fig.~1a).
The value given against each profile is for the peak surface brightness
B$_{\rm [O~III]}$
(uncorrected for interstellar extinction) in units 
of 10$^{-6}$ erg s$^{-1}$ cm$^{-2}$ sr$^{-1}$ \AA$^{-1}$.
}
\label{Fig5} 
\end{figure*}

\end{document}